\documentclass[apj]{emulateapj}
\usepackage{amsmath}
\usepackage{graphicx}
\usepackage{epsfig}

\usepackage{url}
\usepackage{indentfirst}
\begin{document}
\title{On the Polarized Nonthermal Emission from AR~Scorpii}
\author{Takata, J.\altaffilmark{1} and Cheng, K.S.\altaffilmark{2}}
\email{takata@hust.edu.cn, hrspksc@hku.hk}
\altaffiltext{1}{School of Physics, Huazhong University of Science and Technology, Wuhan 430074, China}
\altaffiltext{2}{Department of Physics, The University of Hong Kong, Pokfulam
    Road, Hong Kong}
\begin{abstract}

  We study  linear polarization of optical emission from white dwarf (WD) binary system
  AR~Scorpii. The optical emission from this binary is modulating with
  the beat frequency of the system, and it is
  highly polarized, with the  degree of the polarization  reaching $\sim 40$\%.
  The angle of the polarization monotonically increases with the spin phase,
  and the total swing angle can reach $360^{\circ}$ over one spin phase.
  It is also observed  that the morphology of the pulse profile and the degree of
  linear polarization evolve with the orbital phase. These polarization properties can constrain the
  scenario for  nonthermal emission from AR Scorpii.  In this paper, we study  the  polarization properties predicted
by the emission model, in which (i) the pulsed optical emission is produced by the synchrotron emission
from relativistic electrons trapped by magnetic field lines of the WD and (ii) the
emission is mainly produced at  magnetic mirror points of the electron motion. We find that
this model can reproduce the large swing of the polarization angle,
provided that the  distribution of the initial pitch angle of the electrons that are leaving the M-type star
is biased to a smaller angle rather than a uniform distribution. The observed
direction of the swing suggests that the Earth viewing angle is less than $90^{\circ}$
measured from the WD spin axis. The current model prefers an Earth viewing angle of $50^{\circ}-60^{\circ}$ and
a magnetic inclination angle of  $50^{\circ}-60^{\circ}$ (or $120^{\circ}-130^{\circ}$).
We discuss that
  the different contribution of the emission from M-type star to total emission
    causes  a large variation in the pulsed 
fraction and the degree of the linear polarization  along the orbital phase.

\end{abstract}

\section{Introduction}

AR~Scorpii (hereafter AR Sco) is a binary system composed of a  white  dwarf (WD)
and an M-type main-sequence star, which has a radius of $R_*\sim 0.3R_{\odot}$ and
a mass of $M_*\sim 0.3M_{\odot}$ (\citealt{ma16,bu17}). The WD of AR~Sco shows fast spinning
with a period of $P_{WD}\sim 117$s and an orbital period of $P_{orb}\sim 3.65$hr.
The distance to the system from  Earth is estimated to be  $d\sim 110$pc. AR~Sco is the first WD that shows a
radio emission modulating with the spin period of the WD.
\cite{ma16} measure a spin-down rate of the WD of $\dot{f}\sim -2.86\times 10^{-17}{\rm Hz~s^{-1}}$ and
estimate $B_s\sim 10^8$G for the surface magnetic field of the WD. \cite{po18a}, on the other hand, report that
new optical data are inconsistent with the published spin-down rate but  consistent  with a constant
spin period. Later, \cite{st18} measure a spin-down rate of $\dot{f}\sim -5.14\times 10^{-17}{\rm Hz~s^{-1}}$.
Although more observation would be required to determine the spin down rate, the suggested values indicate
that the radiation from AR~Sco is powered by dissipation of the high magnetic field ($>10^8$G) of the WD.
AR~Sco could  eventually evolve to  a 'polar', for  which a sufficiently large magnetic field  locks
the two stars into synchronous rotation with an orbital period of  $\sim 100-500$minutes (Ferrario et al. 2015).

AR~Sco is  very unique WD binary system with a pulsed emission from the radio to
soft X-ray bands (\citealt{ma16,ta18})  and with a linear polarization of the optical
emission (\citealt{bu17, po18b}). \cite{ma16} report that the radio/optical/UV emission from AR~Sco
is  modulating with a beat frequency ($\nu_b=\nu_s-\nu_o$, where $\nu_s$ and $\nu_o$
are the spin frequency of the WD and orbital frequency, respectively),
and the pulse profile averaged over the orbital phase
shows a  double-peak structure with a phase separation of  $\sim0.5$ in the optical/UV bands and  $<0.5$ in the radio bands. 
\cite{ta18} find that the soft X-ray
emission from AR~Sco also modulates with the beat frequency, and the positions of the
double peak are aligned with the optical/UV peak. This multiwavelength observation
suggests that the same population of electrons produces the observed pulsed
emission from radio to soft X-ray band. The evolution of the  pulse profile of optical/UV emission along the orbital phase is also observed (\citealt{po18b,ta18}). Based on the results of data taken by  the Optical/UV Monitor Telescope on $XMM$-Newton, \cite{ta18} report
that the pulse profile in  optical/UV bands
evolves with the orbital phase, and it changes from the double-peak structure around inferior conjunction of the M-type star's
orbit to a broader peak with no clear double-peak structure at   the superior conjunction.
Since no accretion feature has been discovered in the X-ray emission (\citealt{ma16, ta18}), the broadband emission
likely originates from  the  synchrotron radiation from the relativistic electrons.

It is observed that the WD heats up a half-hemisphere of the M-type star.
Besides the modulation of the beat frequency, the radio, optical, and X-ray emissions of AR~Sco also
show the orbital modulation (\citealt{ma16, ta18, sta18,st18}). The intensity peak of
the emission is observed at around the  superior conjunction of the M-type star's orbit. The spectrum of the  X-ray component
modulating with the orbit is well explained  by the emission from the optically thin thermal plasma  with
several different temperatures \citep{ta18}.  It has been observed that the optical maximum is prior to the superior
conjunction and gradually shifts with time \citep{li17}. \cite{ka17} interprets this orbital shift as a consequence of
either (1)  major magnetic dissipation at the leading surface of the M-type star or (2) the precession of the WD's spin axis.
These observed properties of the orbital modulation will be consistent with the scenario 
that the spin of the M-type star is synchronized  with the orbital motion and the half-hemisphere (day side) of the M-type star is heated by the WD's magnetic field. 
The magnetic dissipation/reconnection process on the M-type star surface will heat
up the plasma to a temperature of several keV and also  accelerate the
electrons to a  relativistic speed. 

Another evidence of the nonthermal emission of  AR~Sco is the discovery of the linear polarization in the
optical bands \citep{bu17}. The polarization of the optical emission from AR~Sco is characterized by a large linear polarization and a large swing  of the polarization angle (P.A.) along  the spin phase ($\phi_s$) of the WD. The polarization degree  can reach  
$\sim 40$\% at pulse peak, and  the P.A. swings $\sim 360$ degree in one spin phase. 
 A recent extensive study found that the pulse profile of the polarized emission
is remarkably stable over time and from  orbit to orbit (\citealt{po18a, po18b}). These polarization features
also support the hypothesis that the pulsed emission  originates from the synchrotron
radiation in the magnetosphere of the WD.
Interestingly, \cite{po18b} measure a  circular polarization, peaking at a value of $\sim 3\%$.
Unlike the optical emission, the radio emission from AR Sco shows a weak linear
polarization ($\le10\%$), but a strong circular polarization that can reach a
the level of $30\%$. \cite{sta18} conclude that the radio
emission of AR~Sco is dominated by cyclotron emission from nonrelativistic particles on the M-type star. 

After the discovery of the pulsed emission from the AR~Sco, several studies have discussed the emission scenario. 
\cite{ge16} discuss that the M-type star interacts with the WD's open magnetic field lines that extend beyond
the light cylinder ($\varpi=P_{WD}c/2\pi=5.6\times 10^{11}{\rm cm}$), and
an electron/positron beam from the WD's polar cap sweeping the stellar wind.
They argue that a
bow shock propagating into the stellar wind accelerates the electrons in the wind. In \cite{ta17},
we argue that the magnetic interaction on the M-type star
surface creates a population of the relativistic electrons, and the closed magnetic field lines of the WD trap
the accelerated electrons by the magnetic mirroring. We propose  that
the synchrotron emission from the first magnetic mirror point after the injection from
the M-type star dominates the observed emission, and the double-peak profile occurs 
  as a result of the contribution of the emission from both the  north and south  poles.
\cite{po18a,po18b} also argued that  an increase in synchrotron emission toward
each magnetic mirror point would  explain the beat  modulation of the optical emission. They suggest that  the  double-peak structure  can be understood as a result of the double-lobed emission profile of
synchrotron beaming from the two poles. 
\cite{be18} discusses the nonthermal emission of AR~Sco with a
hadronic model. The model argues that  the relativistic electrons and hadrons are accelerated in a strongly magnetized
turbulent region around the M-type star and that the primary electrons and/or secondary electron/positron pairs
from decay of pions ($p+p\rightarrow \pi^0+\pi^{\pm}$) produce nonthermal radiation.

\cite{po18b} study the geometry of the synchrotron emission in order
to  explain the polarization characteristic.  They find that the emission site locked
  in the WD rotating frame can reproduce  detailed features of the observed polarization, while
  the emission region fixed in
  the binary frame (e.g. irradiation face of the secondary star) could not.
  They argue  that the emission is around the magnetic mirror point and suggest
  that the polarization modulation
  occurs as a result of an enhanced injection of relativistic electrons into the magnetosphere of the WD when
the magnetic axis points toward the secondary. 

 In this paper, we will apply our emission scenario \citep{ta17} to calculate the polarization characteristic, since the polarization information provides an additional constraint on it. In  \cite{po18b}, a geometrical model is explored  to 
discuss the polarization characteristic. In the present paper, on the other hand, we will calculate
the polarization of the synchrotron radiation by solving dynamics of the injected
electrons.  We will discuss how the polarization characteristic depends on the parameters,
e.g. the initial pitch angle,  system inclination angle, etc. We constrain the direction of the spin axis
of the WD  projected on the sky and the Earth viewing angle from the observed polarization characteristics. 

In section~\ref{model}, we briefly
introduce our emission model and also introduce our method to calculate the linear polarization of the
synchrotron emission. In section~\ref{result}, we summarize our results and show how the angle and degree
of the linear polarization evolve with the spin phase. We  also discuss the dependency of the polarization characteristics on the orbital phase and the viewing
geometry. In section~\ref{disc}, we summarize our results and discuss  the possible viewing geometry inferred from the polarization characteristics. Finally, we argue the difference between the
optical emission processes  of the AR~Sco
and a neutron star pulsar, the Crab (PSR B0531+21).

\section{Model}
\label{model}
\begin{figure}
  \centering
  \epsscale{1}
  \includegraphics[scale=0.5]{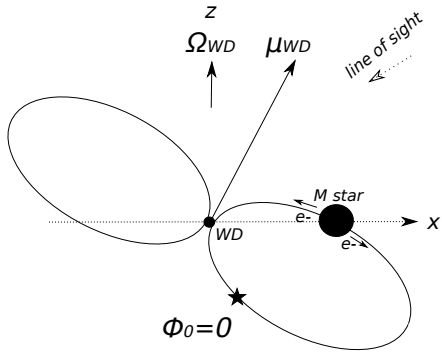}
  \caption{Schematic view of the AR~Sco system and the coordinate. The axes of the WD's spin ($\vec{\Omega}_{WD}$)
    and orbital motion are assumed to be parallel and are defined as the $z$-axis. The $x$-axis is defined so that the observer is located with the plane made by the $x-$ and $z-$axes. We define the orbital phase $\Phi_o=0$ at the inferior conjunction of the M-type star.
    The magnetic axis $\vec{\mu}_{WD}$ is inclined from the spin axis. }
  \label{WD}
\end{figure}
\subsection{Energy injection}
Figure~\ref{WD} illustrates the schematic view of the AR~Sco system explored in this paper. As depicted  in Figure~\ref{WD},
we define the zero orbital phase ($\Phi_o=0$) at the inferior conjunction of the orbit of the M-type star.
We assume that a magnetic dissipation process eventually causes an ablation of the matter (hadrons and electrons)
from the M-type star surface and an acceleration of the electrons to the relativistic speed. We estimate the dissipation rate from (\citealt{lai12,bu17})
\begin{eqnarray}
  L_{B}&=&\frac{B^2}{8\pi}(4\pi R_*^3\delta)\Omega_{WD}
  \sim 2.8\times 10^{32} {\rm erg/s} \nonumber \\
  &\times &\left(\frac{\mu_{WD}}{10^{35}{\rm G~cm^3}}\right)^{2}
  \left(\frac{\delta }{0.01}\right)\left(\frac{R_*}{3\cdot 10^{10}{\rm cm}}\right)^3 \nonumber \\
  &\times& \left(\frac{a}{8\cdot 10^{10}{\rm cm}}\right)^{-6}
  \left(\frac{P_{WD}}{117{\rm s}}\right),
  \label{injection}
\end{eqnarray}
where $\Omega_{WD}=2\pi/P_{WD}$, $\mu_{WD}$ is the WD's magnetic dipole moment,  $a$ is the separation between two stars,
and $\delta\sim 0.01$ is the skin depth on the M-type star surface. We assume
 that  most of the released  energy is used to accelerate the electrons, and the typical  Lorentz factor is estimated to be $\gamma_0\sim L_{B}/(\dot{N}_em_ec^2)$, where $\dot{N}_e$ is the rate of the electron injection in the closed magnetic field line region
 of the WD.  Because of the charge conservation,
 we may assume that the number of injected electrons is equal to the proton number in the
 outflow.  The dissipation of the magnetic
 energy forms the  outflow of the matter from the stellar surface.
 If we denote the efficiency $\chi$ to form the outflow,
 the injection rate of the  protons  may be estimated from 
\begin{eqnarray}
  \dot{N}_p&\sim&\frac{\dot{M}}{m_p}=
  \frac{\chi L_{B}}{\frac{1}{2}m_pv_{esc}^2} \nonumber \\
  &\sim &5\times 10^{40}\chi\left(\frac{L_{B}}{10^{32}{\rm erg/s}}\right)
  \left(\frac{v_{esc}}{5\cdot 10^{7}{\rm cm/s}}\right)^{-2}{\rm /s},
  \label{dotn}
\end{eqnarray}
where $v_{esc}=\sqrt{2GM_*/R_*}$ is the escape velocity.
In the current study, $\chi\sim 10^{-5}$, which indicates that the
typical Lorentz factor of $\gamma_0\sim 50$, will be chosen to fit the observed spectrum.

\subsection{Equation of motion}
With the Lorentz factor of $\gamma_0\sim 50$, the time scale of the synchrotron loss at around the M-type
star is estimated as
\[
\tau_{syn}\sim 400{\rm s}\left(\frac{\mu_{WD}}{10^{35}{\rm G cm^3}}\right)^{-2}
\left(\frac{a}{8\cdot 10^{10}{\rm cm}}\right)^{6}\left(\frac{\gamma_0}{50}\right)^{-1},
\]
which is longer than the crossing time scale of $a/c\sim 2.7{\rm s}$. Hence,
most of the injected electrons migrate to the  inner magnetosphere of the WD.
For the electrons moving along the magnetic field, the evolution of the Lorentz factor,
$\gamma$, and the perpendicular momentum of an electron  may be described by \citep{ha05}
\begin{equation}
  \frac{d\gamma}{dt}=-\frac{P_{\perp}^2}{t_s}
  \end{equation}
and
\begin{equation}
  \frac{d}{dt}\left(\frac{P_{\perp}^2}{B}\right)
  =-2\frac{B}{t_s\gamma}\left(\frac{P_{\perp}^2}{B}\right)^2,
\label{perp}
\end{equation}
where $B$ is the strength of the  local dipole magnetic field, $t_s=3m_e^3c^5/(2e^4B^2)$,  and 
$P_{\perp}=\gamma\beta\sin\theta_p$, with $\beta=v/c$ and $\theta_p$ the pitch angle.
We can see from the equation~(\ref{perp}) that  if the synchrotron cooling time scale $\tau_{s}=\gamma
t_s/P_{\perp}^2$ is longer than the crossing time scale $\tau_{c}=r/c$,
the magnetic mirror for the particle that starts from  $r=a$ will  occur at the point
\begin{equation}
  r_m\sim a\sin\theta_{p,0}^{2/3}, 
\label{rm}
\end{equation}
where  $\theta_{p,0}$ is the initial pitch angle. In this paper, we assume a 
pure magnetic dipole field of the WD, for simplicity. 

For each magnetic field line that sweeps across the M-type star,  the electrons are injected into  both the northern and souther
hemispheres  from the position of the  M-type star with a rate described by equation~(\ref{dotn}).
The injected electrons with an initial pitch angle
  of $\theta_{p,0}>(R/a)^{3/2}$ cannot reach the stellar surface owing to the magnetic mirror effect, and they are  trapped between two mirror points connected by
  the magnetic field line. Such a trapped electron will be eventually absorbed by the M-type star.

  In the current model, the emission from the first magnetic mirror points mainly  contribute to the
  observed  emission, and the peak in the light curve is formed by the emission
  from the electrons
  that are injected when the magnetic pole points toward the companion (for details, see \citealt{ta17}). In Figure~\ref{WD}, for example, the north pole points toward the secondary, and emission from the electron injected into the southern  hemisphere creates one peak. After 0.5 spin phase, the south pole points toward the secondary, and emission
  from the electron injected into the northern hemisphere creates another peak.

\subsection{Radiation process}
\label{radiation}
The observed nonthermal spectrum extends from the radio to the soft X-ray bands.
To explain such a broadband spectrum, we assume that the acceleration process
on the M-type star surface  forms a power-law distribution in the electron's Lorentz factor:
\begin{equation}
  f(\gamma)=K_0\gamma^{-p},~\gamma_{min}\le\gamma\le\gamma_{max},
  \label{distri}
\end{equation}
where $\gamma$ is the Lorentz factor of the electron.  \cite{ta18} fit that broadband spectrum
with a  power-law index of $p=3$. We assume  the minimum Lorentz factor as  $\gamma_{min}\sim \gamma_0$,
and estimate  the maximum Lorentz factor from  the condition that the synchrotron
cooling time scale $t_s\sim 9m_e^3c^5/(4e^4B^2\gamma)$ is equal to the acceleration time scale
of $t_a\sim \gamma m_ec/(\xi eB)$, where $\xi\le 1$ represents the efficiency of the acceleration.
In equation~(\ref{distri}), the normalization $K_0$ is calculated from the condition that
$\int \gamma m_ec^2f(\gamma)d\gamma=L_B$. The power per unit energy
of the  synchrotron radiation for each   electron is calculated from \citep{ry86}
\begin{equation}
  P_{syn}=\frac{\sqrt{3}e^3B\sin\theta_p}{hm_ec^2}F_{syn}\left(\frac{E}{E_{syn}}\right),
\end{equation}
and
\begin{equation}
F_{syn}(x)=x\int_x^{\infty}K_{5/3}(y)dy,
\label{fsyn}
\end{equation}
where $E_{syn}=3he\gamma^2B\sin\theta_p/(4\pi m_ec)$, and $K_{5/3}$ is the modified Bessel function of order 5/3.

\subsection{Polarization}
\begin{figure}
  \centering
  \epsscale{1}
  \includegraphics[scale=0.5]{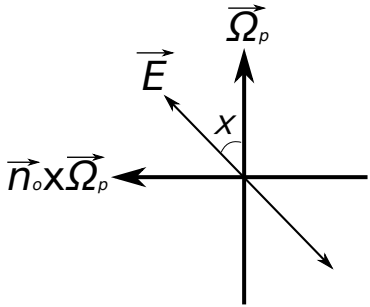}
  \caption{Definition of the angle of the linear polarization. $\vec{\Omega}_p$ and $\vec{n}_o$
    are the direction of the projected spin of the axis seen from the observer and the direction of
    the observer seen from the WD, respectively. The direction of the polarization the P.A. $\chi$ increases in  the counterclockwise,
    when looking at the source.
  }
  \label{goem}
\end{figure}

In this  section, we describe our method to calculate the polarization of the synchrotron radiation from the trapped electrons. We apply  the method developed by \cite{ta07a}, where we calculate
the polarization characteristic of the synchrotron radiation from the electrons/positrons  in the magnetosphere of the Crab pulsar.
The unit vector of the velocity of the electron gyrating around the magnetic field line may be described by
\begin{equation}
  \vec{n}_e=\pm \beta_0\cos\theta_p\vec{b}_{||}+\beta_0\sin\theta_p\vec{b}_{\perp}+\beta_{co}\vec{e}_{\phi},
\label{emd}
\end{equation}
where the vectors $\vec{b}_{||}=\vec{B}/B$, $\vec{b}_{\perp}$, and $\vec{e}_{\phi}$
represent
the unit vectors along the magnetic field line, perpendicular to the magnetic field, and in the azimuthal (corotation with the WD)  direction, respectively.
The plus sign and minus sign in the first term consider  the motion parallel and  antiparallel
to   the magnetic field line, respectively.
The second and third terms represent the gyration motion and the corotation motion with the WD' spin,
respectively.  Since the direction of the gyration motion of the electron is counterclockwise as seen from the direction of
the magnetic field, we may express the unit vector $\vec{b}_{\perp}$ as 
\begin{equation}
  \vec{b}_{\perp}=\cos\delta\phi_{g}\vec{k}+\sin\delta\phi_{g}\vec{b}\times \vec{k},
\label{bperp}
\end{equation}
where $\vec{k}$ is any unit vector perpendicular to the magnetic field line and $\delta\phi_g$ represents the phase of the gyration motion ($0\le\delta\phi_g<2\pi$).   Since the electrons are relativistic, we calculate $\beta_0$ at each position from the
condition that $|\vec{n}_e|=1$ and $\beta_{co}=\varpi/\varpi_{lc}$, where $\varpi_{lc}$ is
the light cylinder radius of the WD. We assume that the emission direction of
the relativistic electron coincides with the direction of the motion given by  the equation~(\ref{emd}).

We assume that after the electrons leave from  the M-type star surface,
they do not gain  energy and  just loose their energy via
the synchrotron radiation. For the synchrotron radiation,
the polarization vector is parallel to the direction of the centripetal acceleration for the gyration motion,
and it is  perpendicular to the direction of the local magnetic field. From equation~(\ref{bperp}), the polarization direction
of the synchrotron radiation may be determined as
\begin{equation}
 \vec{n}_{p}=-\sin\delta\phi_g\vec{k}+\cos\delta\phi_g\vec{b}\times \vec{k}.
\end{equation}

In the calculation, we define the $z$-axis as the spin axis  of the WD, which is assumed to  be parallel to the axis
of the orbital motion (see Figure~\ref{WD}). The $x$-axis is chosen so that  the observer is located at the first quadrant in the ($x, z$) coordinate.
With this coordinate system, the direction of the observer is expressed as
\begin{equation}
  \vec{n}_o=\sin\theta_o\vec{e}_x+\cos\theta_o\vec{e}_z,
  \end{equation}
where $\theta_o$ is the angle of the line of sight measured from the spin axis $\vec{e}_x$ and $\vec{e}_z$ are the unit vectors. 

For each emission point, we consider the emission from all gyration phases, that is, $0\le \delta\phi_g <2\pi$,
and pick up the gyration phase that produces the synchrotron photon propagating toward the Earth.
Then, we calculate the  electric vector of the ``observed''  electromagnetic wave from 
\begin{equation}
  \vec{E}_{em,i}\propto\vec{n}_{p,i}-(\vec{n}_o\cdot\vec{n}_{p,i})\vec{n}_o,
\label{ne}
\end{equation}
where ``$i$'' denotes   each radiation. 
The degree of the linear polarization (hereafter P.D.) is estimated from \citep{ry86}
\begin{equation}
  \Pi_i(E)=\frac{G_{syn}(x)}{F_{syn}(x)},
\end{equation}
where $x=E/E_{syn}$, $F_{syn}$ is given by the equation~(\ref{fsyn}) and $G_{syn}(x)=xK_{2/3}(x),$ with $K_{2/3}$ being
the modified Bessel function of order 2/3.

To calculate the Stokes parameters, $Q_i$ and $U_i$, for each radiation,
we define the polarization  angle  (hereafter P.A.) $\chi_i$ to be the angle
between the polarization vector (\ref{ne}) and the direction of the spin axis projected on the observer sky,
$\vec{\Omega_p}=\vec{\Omega}_{WD}-(\vec{n}_o\cdot\vec{\Omega}_{WD})\vec{n}_o$ (Figure~\ref{goem}),
\begin{equation}
  \cos\chi_i=\frac{\vec{E}_{em,i}\cdot\vec{\Omega}_p}{|\vec{E}_{em,i}||\vec{\Omega}_p|}. 
\end{equation}
For the conventional definition,  the P.A. $\chi_{i}$ increases in counterclockwise, when looking at the source.
The Stokes parameters for each radiation are 
$Q_i(E)=\Pi_i(E)I_i(E)\cos2\chi_i$ and $U_i(E)=\Pi_i(E)I_i(E)\sin2\chi_i$, where $I_i(E)$ is the intensity.
By collecting the observed radiation at each bin of the spin phase, $I(\phi_s)=\sum I_i$,
$Q(\phi_s)=\sum Q_i$, and $U(\phi_s)=\sum U_i$, we calculate the P.D. and P.A. for each bin from
\begin{equation}
P(\phi_s)=\frac{\sqrt{Q^2(\phi_s)+U^2(\phi_s)}}{I(\phi_s)}, 
\end{equation}
and
\begin{equation}
  \chi(\phi_s)=\frac{1}{2}{\rm tan}^{-1}\left[\frac{Q(\phi_s)}{U(\phi_s)}\right],
\end{equation}
respectively.

The  current method cannot apply to estimate
  the circular polarization, since we consider
only emission  in the direction of the particle motion (see equation~(\ref{emd})).
The synchrotron emission  is in general elliptically polarized
if the viewing angle deviates from  the direction of the particle motion.  For a power-law distribution of electrons, for example,
the degree of the circular
polarization of the synchrotron emission from the electrons with
Lorentz factor $\gamma_e$ is given by (\citealt{sa72,na16})
\begin{equation}
P_{circ}\sim -\frac{P_{lin}}{\gamma_e}\left(\cot\psi+\frac{1}{p+2}\frac{1}{Y(\psi)}\frac{dY(\psi)}{d\psi}\right),
\end{equation}
where $P_{lin}$ is the degree of linear polarization, $\psi$ is the angle between the particle motion and the emission,
and $Y(\psi)$ is the pitch-angle distribution. The ratio between circular and linear polarization is of the order of
$1/\gamma_e$. \cite{po18b}, on the other hand, measure the circular polarization, peaking at a value of $\sim 3\%$. Such a high level
of the circular polarization could imply a high level of pitch-angle anisotropy.   A detailed calculation
for the circular polarization is more complicated than that for the linear polarization since
we have to take into account the emission in all directions relative to  the particle motion and the pitch-angle distribution.
We  therefore focus on the linear polarization in this paper and will discuss  the circular
polarization in a subsequent study. 

  \subsection{Pitch-angle distribution}
Besides the energy distribution of the injected electrons, we also consider the effect of the distribution of
the pitch angle. Within the current framework of the emission model,
we will see in section~\ref{result} that the
pitch-angle distribution sensitively affects the polarization characteristic. The distribution of
the initial pitch angle of the relativistic electrons leaving the M-type star may be related to the acceleration
process. For example, the standard acceleration model of the  neutron star pulsar assumes that 
the electric field along the magnetic field line produces a relativistic particle. This acceleration process
produces a population of the relativist particles with a small pitch angle. For the shock acceleration (e.g.
in the magnetic turbulent), on the other hand, the accelerated particles would have various pitch angles  (\citealt{ac01,ka16}). In this study, we explore the pitch-angle distribution with  a function form of 
\begin{equation}
  \frac{d\dot{N}_e}{d\cos\theta_{p,0}}=C_0+C_1\frac{  {\rm exp}\left[-\frac{(\theta_{p,0}-\theta_0)^2}{2\sigma_0^2}\right]}{A},\
  ~~0<\theta_{p,0}<\pi/2,
  \label{pdis}
\end{equation}
where $A=\int_0^{\pi/2}{\rm exp}\left[-(\theta_{p,0}-\theta_0)^2/(2\sigma_0^2)\right]{\rm d}\cos\theta_{p,0}$. With the above equation,
we assume that the initial pitch-angle distribution is composed of the isotropic component plus  the
anisotropic component described by the Gaussian-like function.
In this study, the ratio of  the constant factor  $C_0/C_1$, the central value $\theta_0$, and the dispersion $\sigma_0$ are
model-free parameters.  

    \section{Results}    
\label{result}
    \begin{figure*}
  \centering
  \epsscale{1}
  \plotone{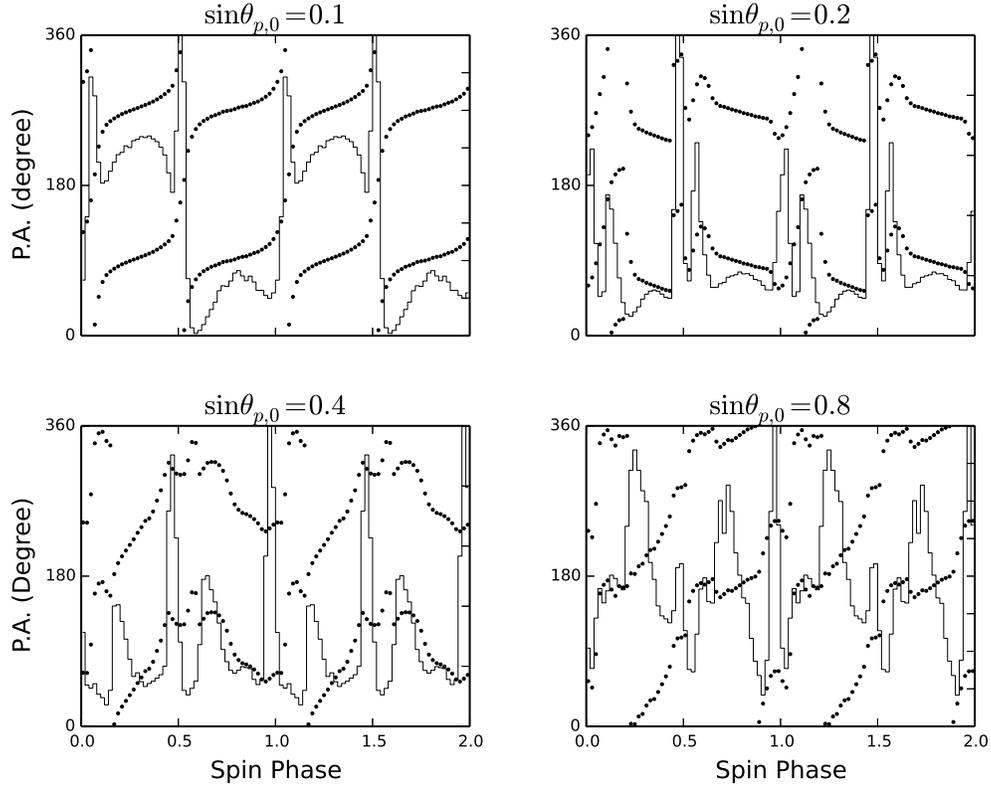}
  \caption{Evolution of the polarization angle (filled circles) with  spin phase
    for a specific initial pitch angle.
    The pulse profile for each case  
    is plotted with the solid histogram. The results are calculated with $\Phi_o=0.5$,
    $\alpha=60^{\circ}$, and $\zeta=65^{\circ}$, respectively.}
  \label{pitch}
\end{figure*}

\begin{figure*}
  \centering
  \epsscale{1.2}
  \plotone{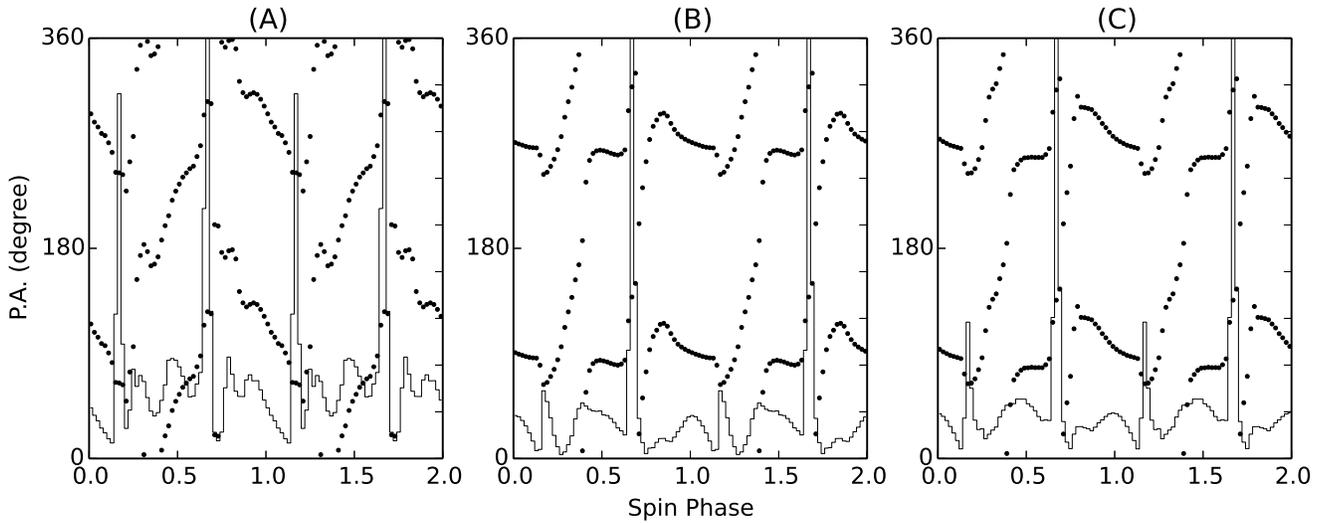}
  \caption{Dependency of the polarization angle (filled circles)  on the distribution of the initial
    pitch angle. (A) : Uniform distribution between $0\le \theta_{p,0}\le \pi/2$. (B) : Gaussian-like distribution with $\theta_{0}=0.2$rad and $\sigma_{0}$=0.2rad. (C) Distribution created by
    uniform component
    plus Gaussian-like component with  $\theta_{0}=0.2$rad and $\sigma_{0}$=0.2rad and $C_1/C_0=1$ in equation~(\ref{pdis}). The results are calculated with $\alpha=60^{\circ}$, $\zeta=60^{\circ}$, and  $\Phi_o=0.5$.}
  \label{pangle}
\end{figure*}

\begin{figure}
  \centering
  \epsscale{1.2}
    \plotone{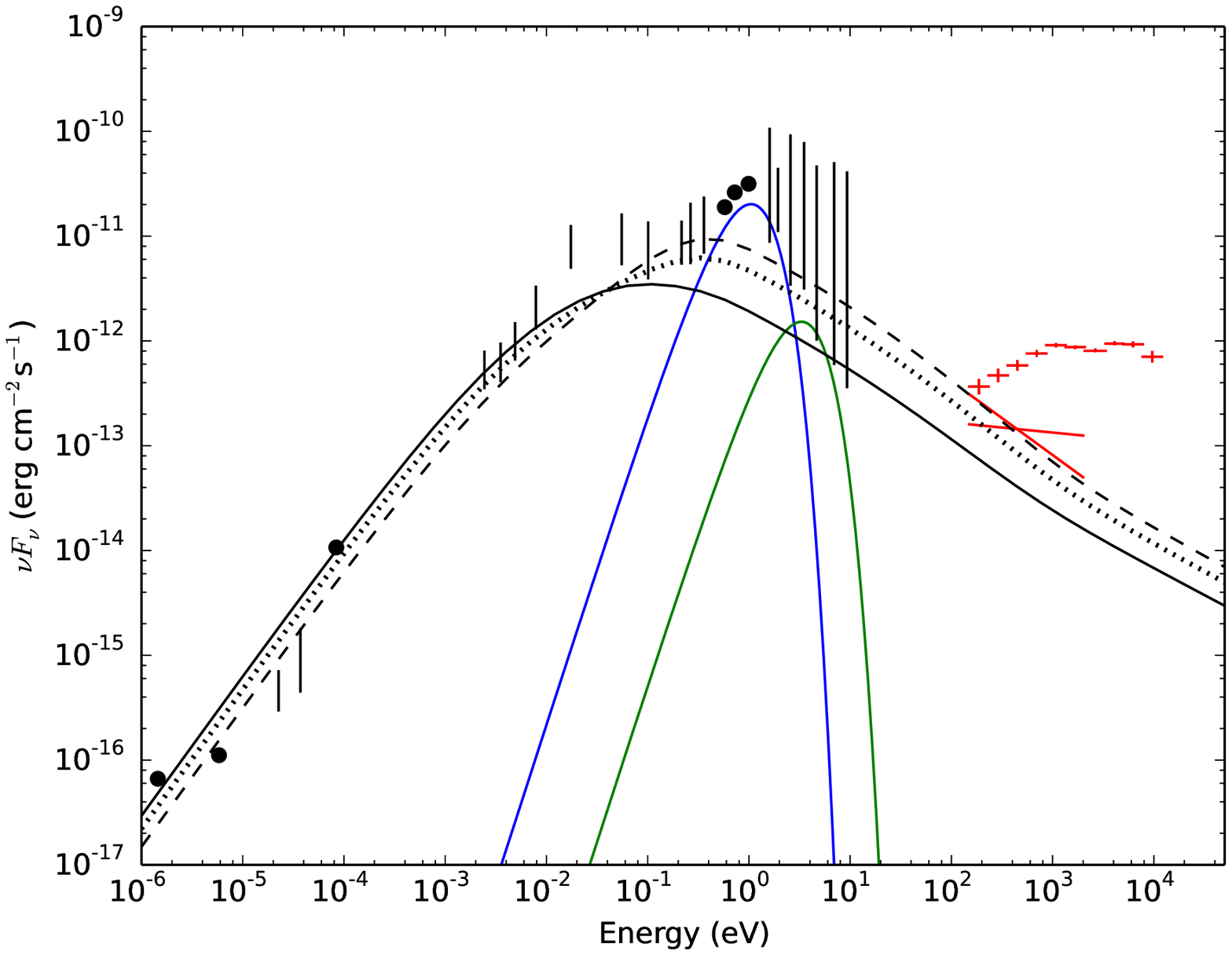}
    \caption{Spectral energy distribution of AR~Sco. The black solid, dashed, and dotted lines are the calculated spectrum with the initial pitch angle distributions for the panels (A), (B) and (C) in Figure~\ref{pangle}, respectively. The data are taken from Takata et al. (2018) for X-ray bands and Marsh et al. (2016) for the radio/optical/UV bands. The red lines and plus signs represent the pulsed and nonpulsed components, respectively, 
            of the X-ray emission.
            The blue and the green lines are the models for the  blackbody emission from the M-type star and the WD, respectively.
            In the model calculation, the initial distribution in the Lorentz factor is assumed to be $p=3$. }
\label{sed}
\end{figure}
The size of the M-type star ($R_*\sim 0.3R_{\odot})$ is not negligible compared to the size of the system ($a\sim R_{\odot}$). In the calculation, therefore,
we inject the particles on the magnetic dipole field lines penetrating the day side of the M-type star (a half-hemisphere), and assume that the injection rate is the same for all magnetic field lines.
In the calculation, we assume the inclination angle, $\alpha$, and the viewing angle, $\zeta$,
measured from the spin axis of the WD. We will present the results for Earth viewing angle $\zeta<90^{\circ}$,
because  the observed swing direction of the P.A. predicts it, as discussed in section~\ref{disc}.  The polarization characteristics predicted by
$\alpha=\alpha_0$ and  $\alpha=180^{\circ}-\alpha_0$ are identical from each other.
In Figures~\ref{pitch}, \ref{pangle}, and \ref{orbit}-\ref{zeta}, the evolution of the P.A.
(filled circles) along  the spin phase  is presented with the pulse profile (solid histograms)

\subsection{P.A. and initial pitch angle}
\label{pa}
First, we examine the pulse profile and the swing of the P.A. predicted by a specific initial pitch angle, $\theta_{p,0}$.
Figure~\ref{pitch} summarizes the evolution of the P.A. (filled circles)  along the spin phase of the WD for  the initial pitch angles of $\sin\theta_{p,0}=0.1$ (top left),
0.2 (top right), 0.4 (bottom left) and 0.8 (bottom right). The results are for the initial Lorentz factor $\gamma_0=100$, with which the electrons mainly produce
the optical emission with the synchrotron radiation process.  In each panel, the predicted pulse profile (solid histogram) is also plotted.  

We find in Figure~\ref{pitch} that the calculated P.A. shows a large swing along the spin phase. For the pitch angle $\sin\theta_{p,0}=0.1$, for example,
the pulse profile shows the double-peak structure and the P.A. swings $\sim 180^{\circ}$ at each peak. The  total swing of the P.A. over  one spin period is
$\sim 360^{\circ}$, which can explain  the observed swing angle. For a larger initial pitch angle, the calculated pulse profile is
more complicated, with more than two peaks.  This dependency of the pulse profile and the polarization characteristic are
related to the contribution of the emission from the second and subsequent
magnetic mirror points and the emission from both hemispheres. Equation~(\ref{rm}) shows that the magnetic mirror occurs at the inner magnetosphere for the smaller pitch angle.
This implies that a higher  fraction of the initial energy of the particles is lost
by the synchrotron radiation at the first mirror point. For $\sin\theta_{p,0}=0.1$
in Figure~\ref{pitch}, for example,  most of the initial particle energy is lost at the first mirror point, and the contribution of the
emission from the subsequent mirror points makes small broad  bumps between two main peaks. Since the photons detected
at each spin phase bin are contributed by  a smaller region in the magnetosphere, the angle of the linear polarization shows  a
monotonic increment  along  the spin phase.

For a larger initial pitch angle, since the magnetic mirroring occurs at a smaller magnetic field region (outer magnetosphere), a smaller fraction  of the initial energy is released  as the synchrotron radiation at the first mirror point. Hence, the emission from the subsequent mirror points also creates  noticeable peaks in the calculated pulse profile, as
we can see in Figure~\ref{pitch}. A wider region in the magnetosphere contributes to the emission of each spin phase,
and therefore the resultant  evolution of  the P.A. along
the spin phase and the pulse profile  become more complicated.

\subsection{Pitch-angle distribution}
Figure~\ref{pangle} summarizes the dependency of the polarization  on the distribution of the initial pitch angle, by assuming
the inclination angle $\alpha=60^{\circ}$ and the Earth viewing angle $\zeta=60^{\circ}$. The results are
for the superior conjunction ($\Phi_o=0.5$). In the figure we can see that the predicted P.A. shows a large
swing with the spin phase, but its evolution depends on the initial distribution.
In panel (A), we assume a uniform distribution in $0\le\theta_{p,0}\le \pi/2$,
that is, $C_1=0$ or $\sigma_0=\infty$ in equation~(\ref{pdis}), and we can see  a complicated
evolution of the P.A. with the spin phase. This is because
the emission from high-order magnetic mirror points cannot be ignored in the pulse profiles.
In the panel (B), we consider only the Gaussian component and assume that initial pitch
angles concentrate at $\theta_0=0.2$~rad with a width of $\sigma_0=0.2$~rad. In such a case, more  injected electrons lose their energy at the first magnetic mirror point, and the resultant polarization tends to  increase with the spin phase.  We can expect that as we increase the magnitude of $\sigma_0$, the calculated P.A. and pulse profile shift  to those of the panel~(A).

By comparing the calculated pulse profiles (solid histograms) in the panel (A) and panel (B), we find that  the double-peak structure
is more clearly shown in the pulse profile with the  uniform distribution. In current model, a strong peak is created by each hemisphere, and the double peak structure is produced as a result of the contribution
of the emission from both hemisphere. When the initial pitch angle concentrates to a  smaller value, the emission
from a half hemisphere could be missed by the observer, and the height of one peak is much lower  than other peak, as the panel (B) indicates. To explain the observed double peak pulse profiles, therefore, a uniform component may be  necessary.  

In panel (C) of Figure~\ref{pangle}, we assume that the initial pitch angle is described by the uniform component
plus the Gaussian-like component, and we present the result with $C_1/C_0=1$, that is, we assume that  the total particle
number of the Gaussian component is equal to  that of the uniform component. For the Gaussian component, we assume $\theta_0=0.2$~rad and $\sigma_0=0.2$~rad. In the figure, we can see that the pulse profile and evolution of the
polarization angle are  intermediate between those in the  panels of (A) and (B). For example, we can see that the pulse height of the second peak (minor peak)
is the lower than that in panel (A), but it is higher than that in panel (B). The evolution of the polarization angle with the spin phase is not  as smooth as
panel (B). We find that with current
parameters of $\alpha=60^{\circ}$ and $\zeta=60^{\circ}$,  $360^{\circ}$
of the P.A. swing along the spin phase can be reproduced with $C_1/C_0\ge1$.  Our model therefore suggests that the particle acceleration on
the M-type star surface produces more relativistic electrons having smaller
pitch angle.  As a conclusion of this section, the two component model (panel C) of the initial pitch-angle
distribution explains better both  observed double peak  structure  and polarization characteristics.

The predicted anisotropy of the initial pitch angle distribution
  might be consistent with the observed circular polarization, peaking at a value of $\sim 3\%$ \citep{po18b}.
  Such a high level of the circular polarization would require  a high level of pitch-angle anisotropy. The calculation of the circular polarization based on the current model will be done in a subsequent study.

Figure~\ref{sed} compares the model spectra for the panels (A) (solid line), (B) (dashed line), and
(C) (dotted line) in Figure~\ref{pangle}. In the calculation, we assume $p=3$ for
the power-law distribution in the initial Lorentz factor to explain the observed spectra of the
pulsed emission in optical/UV and soft X-ray bands. In this figure, we can see that the predicted spectrum
becomes harder if the initial pitch-angle distribution is biased in the smaller angle. This can be understood because the mirror point for an electron with a smaller initial pitch angle is closer to the
stellar surface, and therefore the synchrotron emission at the mirror point is harder.
As we can see in the figure, the observed spectra cannot tightly constrain the initial pitch-angle distribution.

\subsection{Evolution with orbital phase}

\begin{figure*}
  \centering
  \epsscale{1}
  \plotone{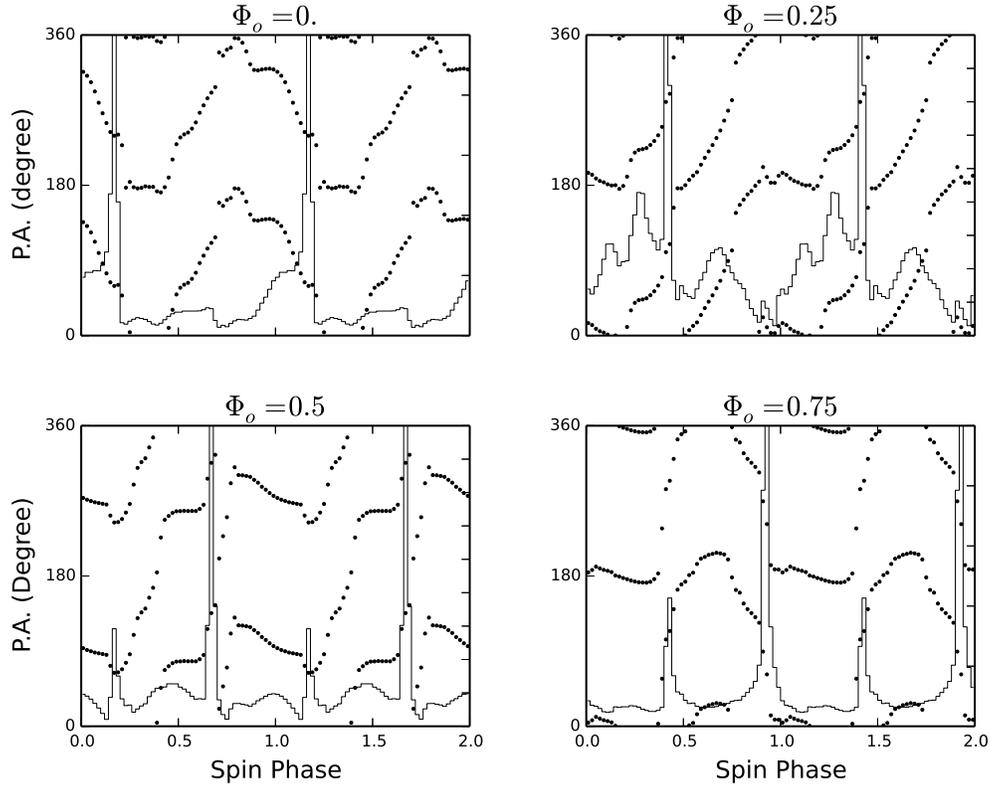}
  \caption{Evolution of the P.A. swing (circles) and the pulse profile (histograms)
    with the orbital phase. The magnetic inclination angle and viewing angle
    are $\alpha=60^{\circ}$ and $\zeta=60^{\circ}$, respectively. The phase $\Phi_0=0$ corresponds to the inferior conjunction of
    the M-type star, and $\phi_s=0$ in all panels  corresponds the spin phase at which the magnetic axis is directed  toward the Earth.
    The results are for $C_0=C_1$ with $\theta_{0}=0.2$ and $\sigma_{0}=0.2$.}
  \label{orbit}
\end{figure*}

Figure~\ref{orbit} summarizes the dependency of the evolution
of the linear polarization (filled circles) and the pulse profiles (histograms) on the orbital phase.
For the pulse profile, we  can see that the position of the peak shifts
with the orbital phase. In the current model, the pulse peaks are mainly created by the emission
of the electrons injected when the magnetic axis of the WD points toward the M-type star \citep{ta17}, and hence
the pulse peak shifts with the orbital motion of the M-type star.

We  see in the figure that the polarization characteristics and the pulse profile depend on  the orbital phase. At the orbital phase $\Phi_o=0.25$ (descending  node of the orbit of the M-type star) and 0.5 (superior conjunction), for example, the P.A.  tends to  increase with the spin phase, while at $\Phi_o=0$ (inferior conjunction)
and 0.75 (ascending node), the swing of the P.A. changes its direction at spin phase $0.6-0.7$.
For $\Phi_o=0$, we find a small second peak in the calculated pulse profile.
Since the M-type star is between Earth and the WD,  the electrons leaving from the M-type star
move away from  Earth. As a result, the strong emission from one hemisphere may  not point toward  Earth. 

We note that the predicted  characteristics of the P.A. and the pulse profiles at $\Phi_o=0.25$ and $\Phi_o=0.75$
are different from  each other; nevertheless, the positions of the M-type star are symmetric
relatively to the axis made by Earth and the WD. The difference originates from the direction of
the WD's spin. In the  current study, we anticipate that the direction of the WD's spin is parallel to the axis of the orbital
motion.  At $\Phi_o=0.25$, since the WD's magnetic field sweeps across the M-type star  from the Earth side to the backside, the trapped electrons initially  move away from the Earth. This makes a pulse profile broader. 
At $\Phi_o=0.75$, the magnetic field sweeps across the M-type star  from the backside to the Earth side, and therefore
the trapped electrons move initially toward the Earth. The emitted photons and the emitting electrons that continuously
produce the photons  may initially move in the same direction. This tends to shorten the difference in the arrival times
of the different photons and tends to make a sharper peak. 

We mention that the current calculation has some discrepancies from the observations. 
  For example, the observed optical/UV pulse profile evolves with the orbital phase. With OM data of the $XMM$-Newton telescope,  we (Takata et al. 2018) show that the shape of the pulse profile rapidly changes at around $\Phi_o=0$ and a small second  peak structure is found in the pulse profile at around
  $\Phi_o=0.1$. This feature is similar to the current model. On the other hand,
  the observed pulse profile at  $\Phi_0=0.5-0.6$ is described by  a broad pulse 
  profile with no clear second peak,
  while the current pulse profile shows a double-peak structure.  For the optical polarization,
\cite{po18b} show that the  total swing of the P.A. over the
spin phase is $360^{\circ}$ and does not depend on the orbital phase.
As we can see in Figure~\ref{orbit}, the current calculation predicts
that the total  swing of  the P.A.  at  $\Phi_O=0$ and $0.75$ phases
is less than $360^{\circ}$. A fine-tuning for parameters (e.g. initial Lorentz factor and pitch-angle distribution and system geometry) will be required to obtain a total $360^{\circ}$ swing
for the whole orbital phase.  This difference between the model and observation
could also indicate that the realistic
geometry and structure of the magnetosphere (e.g. magnetic field structure)
are  more complicated comparing to
current simple treatment. For example, the dipole field approximation could be a very rough treatment owing to the interaction with the companion star.

\begin{figure*}
  \centering
  \epsscale{1}
  \plotone{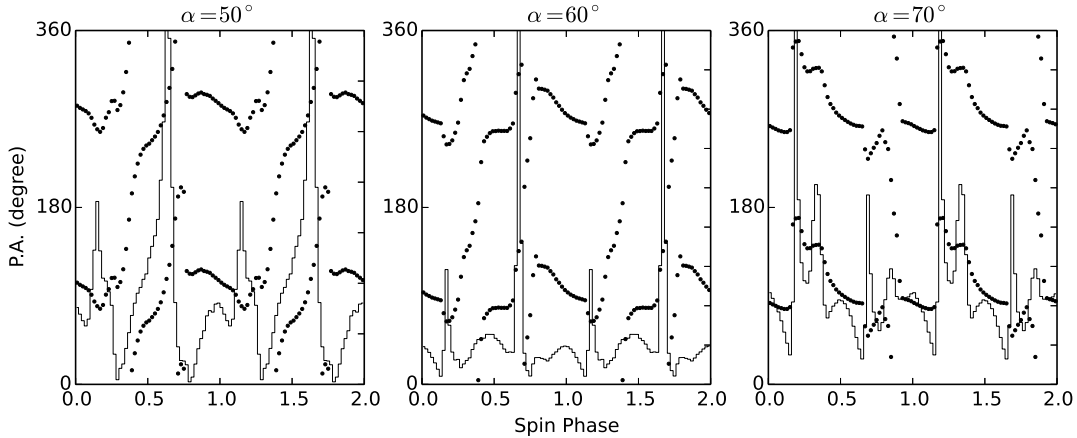}
  \caption{Dependency of the P.A. swing and the pulse profiles   on the magnetic inclination angle. The results  are for $\Phi_o=0.5$,  the Earth viewing angle $\zeta=60^{\circ}$, and $C_0=C_1$.}
  \label{alpha}
\end{figure*}

\begin{figure*}
  \centering
  \epsscale{1}
  \plotone{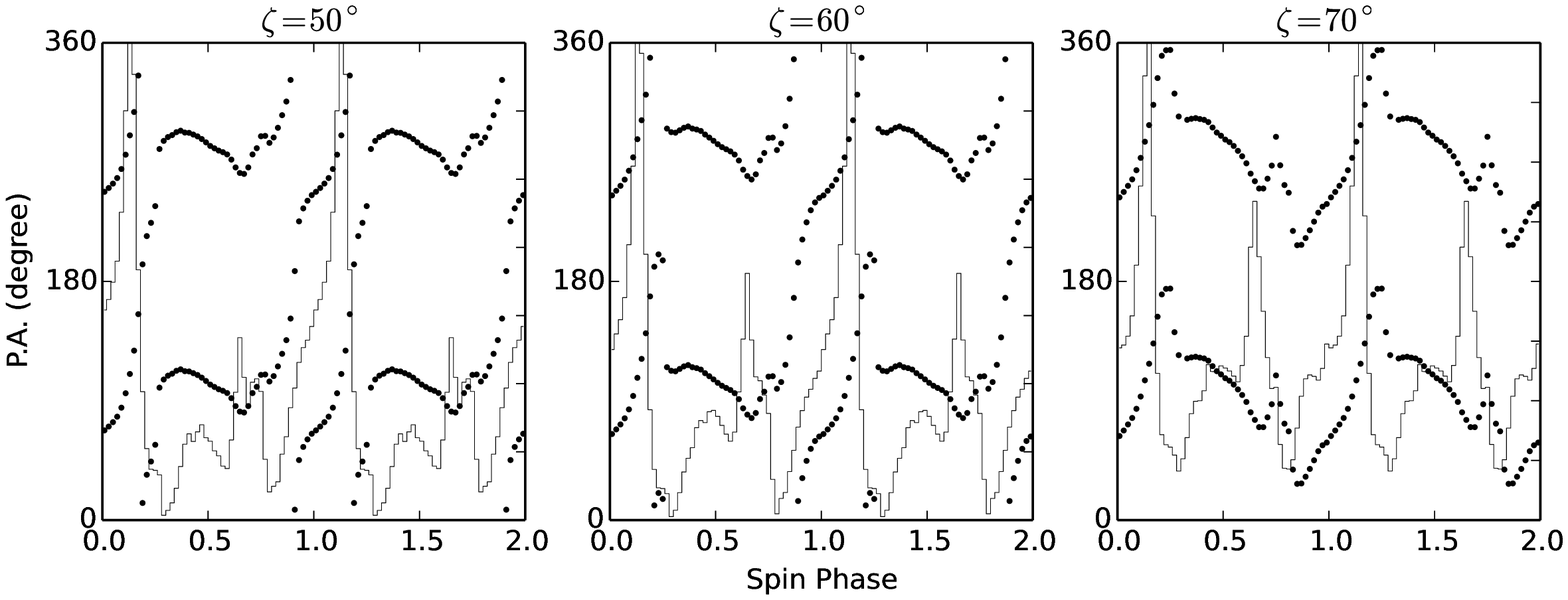}
  \caption{Dependency of the P.A. swing and the pulse profiles on the Earth viewing angle.
    The results  are for $\Phi_o=0.5$, the magnetic inclination angle $\alpha=50^{\circ}$, and $C_0=C_1$.}
  \label{zeta}
  \end{figure*}

\subsection{Dependency on the geometry}
Within the current framework of the model, we find that the calculated polarization characteristic depends on the viewing geometry. For example,
Figures~\ref{alpha} and \ref{zeta} summarize the dependency of the inclination angle of the magnetic axis of the WD and the Earth
viewing angle, respectively. Using  the orbital phase $\Phi_o=0.5$, 
the Earth viewing angle $\zeta=60^{\circ}$, and $C_0=C_1$, 
Figure~\ref{alpha} compares
the results for $\alpha=50^{\circ}$ (left panel), $60^{\circ}$ (middle panel),  and $70^{\circ}$ (right panel). In the figure, we can see  the  tendency of an increase in the  P.A. along the spin phase for  inclination angles $\alpha=50^{\circ}$ and $60^{\circ}$,
while we see a complicated evolution for $\alpha=70^{\circ}$. For
$\alpha=70^{\circ}$, we can see  many sharp peaks in the calculated pulse profile. This shows that with a specific Earth viewing
angle, the emission from the second and subsequent mirror points also contributes to the observed emission and produces a complicated evolution of the P.A. swing. We find therefore that the  current model prefers the magnetic inclination
angle of $\alpha=50^{\circ}-60^{\circ}$ to produce a $360^{\circ}$  swing of the P.A. along the spin phase.
For a  nearly aligned rotator, on the other hand, the model does not
predict the double-peak structure, as discussed in Takata et al. (2017).

In Figure~\ref{zeta}, we summarize the dependency of  the viewing angle on the polarization characteristics with $\alpha=50^{\circ}$,
$\Phi_o=0.5$, and $C_0=C_1$. With these parameters, we can see in the figure that a smaller viewing angle ($\zeta=50^{\circ}$ and $\zeta=60^{\circ}$) shows a monotonic increase in the P.A. with the spin phase, and for  a larger viewing angle (right panel for  $\zeta=70^{\circ}$),
the P.A. changes its swing direction at a spin phase. This dependency on the viewing angle also comes from the dependency of the contribution
from  both hemispheres. From the pulse profiles (histograms in the figure), we can see that the height of the
second peak at $\phi_s\sim 0.7$ increases as the Earth viewing  angle increases.  This is related to the fact that when we look at the system from an  angle closer to the
equator, the difference in the contribution of the emission from the two  hemispheres
decreases  (for the observer with $\zeta=90^{\circ}$, the contributions from two hemispheres are even).  This makes the evolution of the P.A. swing and the pulse profile more complicated. For viewing angle $\zeta\sim 0^{\circ}$, on
the other hand, we expect  no or smaller modulation of the observed
flux with the spin/beat phase. As a result, the  current model prefers  the Earth viewing angle, $\zeta=50^{\circ}-60^{\circ}$,
as well as the magnetic inclination angle $\alpha=50^{\circ}-60^{\circ}$.

\subsection{Degree of linear polarization}
The observed optical emission will be composed of two components:  (1) thermal emission from the M-type star surface and (2) nonthermal emission from the relativistic particles, as Figure~\ref{sed} indicates. It has been observed that the pulsed fraction, which is defined
by the equation ($f_{max}-f_{min})/(f_{max}+f_{min})$, with $f_{max}$ and $f_{min}$ being  the maximum and minimum counts in the
pulse profile, depends on the photon energy bands and on the orbital phase; it becomes
maximum around the inferior conjunction (hereafter INFC) of the M-type star's orbit and
minimum around the superior conjunction (hereafter SUPC). For example, the pulsed fraction
of the UV emission is $\sim 70$\% at the INFC and about $\sim 40$\% at the SUPC \citep{ta18}. Moreover, it has been observed
that the P.D. of the optical emission also depends on the orbital phase and may be related to the pulsed fraction.
\cite{bu17} report that the  the P.D.
can reach at the level of $\sim 30-40$\% at the orbital
phase $\Phi_{ob}=0-0.2$,  while it  is $10-20$\% at $\Phi_{ob}\sim 0.4-0.6$.
These evolutions of the pulsed fraction and P.D. would reflect
(i) an evolution of the linear counts  (synchrotron emission) and/or (ii) an evolution of  contribution of the emission from the day
side of the M-type star.

\begin{figure}
  \centering
  \epsscale{1}
  \plotone{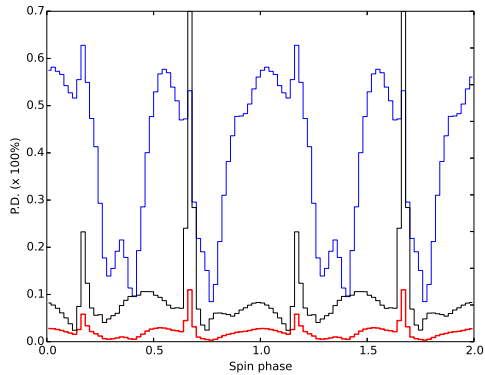}
  \caption{Typical P.D.. The blue line shows the result without 
    the contribution
    of the emission from the M-type star, for which the unpolarized emission is assumed.  The red line considers the contribution of 
    the unpolarized emission to reduce the pulsed fraction to 10\%.
    The results are for $\Phi_o=0.5$, $\alpha=60^{\circ}$,
  $\zeta=60^{\circ}$, and $C_0=C_1=1$.}
    \label{pdeg}
\end{figure}

Figure~\ref{pdeg} shows typical P.D. predicted by the  current  model.
If we consider only the synchrotron emission (blue line), the pulsed fraction
is $\sim 80$\% and the P.D.  can reach $\sim 60$\%, which is higher
  than the observed one,  P.D.$\le 40$\%.  We therefore expect that
  the contribution of the unpolarized emission from the star reduces the P.D.
  To examine the effect of the thermal emission on the P.D.,
  we assume that the thermal emission from the M-type stellar surface is unpolarized. Then, we add the unpolarized emission as
  a DC emission to reduce the pulsed fraction  to $10$\%.
  The calculated P.D. (red line in the figure) is
  significantly reduced to $\le 10$\% and has a peak at the pulse peak,
  which is roughly consistent with the observation around $\Phi_o\sim 0.5$.

  From  \cite{bu17} and \cite{po18b}, for example, we can see that the linear counts averaged over the spin phase are almost constant between 0.1 and 0.5 orbital phase. However, the maximum P.D. of $\sim 40$\%  at around 0.1-0.2 orbital phase is higher than $\sim 20$\% at around 0.4-0.5 orbital phase.
  This difference  in the P.D. would be attributed by the different contributions of the
  stellar emission to the total emission.
  Since the contribution of the emission from the M-type star to the total emission at
  0.1-0.2 orbital phase is smaller than that at 0.4-0.5 orbital phase, the P.D.
  at 0.1-0.2 orbital phase is larger than that at 0.4-0.5 orbital phase.

\section{Discussion and Summary}
\label{disc}
The swing direction of the observed  linear polarization will be able to constrain the viewing geometry. In the  previous
section, we assumed that the viewing angle is smaller  than $\zeta<90^{\circ}$ measured from the
WD's spin axis. This condition is necessary to explain the swing direction of the observed P.A. with
the spin phase,  that is,
counterclockwise. For the viewing angle with  $\zeta>90^{\circ}$, the expected direction is clockwise.
The difference in the swing directions for the two viewing angles  mainly comes from the difference in the directions of the magnetic field line projected on the sky.
We note that
for a specific viewing angle, the inclination angles $\alpha=\alpha_0$ and $\alpha=180^{\circ}-\alpha_0$
do not make  any difference in the morphology of the  pulse profile and the polarization characteristics, since the directions of the magnetic field lines projected on the sky for the  two inclination angles
are identical. Hence, the required condition that  $\zeta<90^{\circ}$
does not depend on whether the magnetic inclination angle is larger or smaller than $90^{\circ}$. 

In Figure~3 of \cite{bu17}, we can see the spin phase where  the swing angle of the observed P.A.  is small  
(e.g. $\phi_s=0.1-0.5$ in the right panel of the figure). In our results, a similar trend can be seen in
the calculated evolution of the P.A. In  Figure~\ref{orbit}, for example, we find a spin phase where
the direction of the linear polarization is almost constant at P.A.$\sim 90^{\circ}$ for $\Phi_o=0.5$ and at P.A.$\sim 0^{\circ}$ (and $180^{\circ}$) for $\Phi_o$=0, 0.25, and 0.75.
Figure~\ref{alpha} and \ref{zeta} also suggest that  the P.A. of the flat region does not depend on the inclination angle
and viewing angle.   In  our definition, the direction of the P.A.$=0^{\circ}$ (or $90^{\circ}$)
is parallel (or perpendicular)
to the direction of the projected spin axis. Therefore, a detailed comparison between the observed and calculated
P.A. may enable us to determine the direction of the spin axis of the WD projected on the sky. 

The double-peak structure of the pulse profile, the large P.D., and the large swing of the polarization angle of the optical
emission from AR~Sco  resemble  those  of the Crab pulsar \citep{ka05}.
Based on our emission model, however, we would like to say that
although  the emission process is the  synchrotron radiation for the two cases,
the particle acceleration/emission regions are  very different from each
other.  It has usually  been argued that the nonthermal
emission of the neutron star pulsar originates from  the
open field line region, for which the magnetic field lines extend beyond
the light cylinder that is defined by the axial distance $\varpi_{lc}=P_sc/2\pi$ ($\sim 1.5\times 10^8{\rm cm}$
for the Crab and $5.6\times 10^{11}{\rm cm}$
for the WD in AR~Sco), and the emission  takes place around the light cylinder.
For the AR~Sco, the M-type star orbits at $a\sim 8\times 10^{10}{\rm cm}$ 
inside the light cylinder of the WD, and hence the emission probably originates
from the closed magnetic field line region, as we have discussed in this paper.
For the Crab pulsar, the pair-creation process of the GeV gamma-rays creates
the electrons and positrons that emit the synchrotron photons.
(\citealt{ta07a,ta07b}).
These secondary pairs will have a small pitch angle of $\sin\theta_p\sim 0.1$, and the emission from them  covers
a part of the sky. A special
relativistic effect (e.g. flight time and aberration) makes a peak in the
observed pulse profile of the Crab pulsar. In the current model of the AR~Sco, the mirror point of the closed magnetic field is the main emission region, and
the special relativistic effect is less important to form a peak
in the pulse profile. 

In summary, we have studied the linear polarization of the nonthermal optical
emission from AR~Sco with the  model developed by  \cite{ta17}. In the model,
the relativistic electrons trapped by the closed magnetic field line region produce the
nonthermal emission with the synchrotron process.  We have found  that the calculated
linear polarization can have  a large swing through the spin phase, although
the evolution of the P.A. with the spin phase depends on various factors (e.g. orbital phase, initial pitch-angle
distribution, and viewing geometries).
The total swing of the observed P.A. over the spin period  can be $360^{\circ}$.
To explain the large swing angle and the double-peak structure of the pulse profile,
 the current model suggests  (1) an isotropy of the initial pitch-angle distribution, which is
biased  to a smaller value, and (2) a moderate
magnetic inclination angle $(\alpha\sim 50^{\circ}-60^{\circ}$) and the Earth viewing angle
$(\zeta\sim 50^{\circ}-60^{\circ})$. We have also shown that the P.D. can reach to $\sim 60$\%, if the emission from the relativistic electrons dominates the emission from the stellar surface.
The different contribution of the emission from the  M-type star on the observed optical
emission will explain the  evolution of the observed P.D. with the orbital
phase. We have discussed
that the origin of the nonthermal emission of the AR~Sco is different from the neutron-star-pulsar-like emission process.
However, AR~Sco is only a sample of the pulsed nonthermal emission from the magnetic WD. More samples will be necessary  to understand the nonthermal nature of the magnetic WDs and the similarity/dissimilarity with the non-thermal emission from neutron star pulsars.

We express our appreciation to an anonymous referee for useful comments and suggestions.
J.T. is supported by NSFC grants of the Chinese Government under 11573010,  11661161010, U1631103 and U1838102.
K.S.C. is supported by GRF grant under 17302315.

\end{document}